\documentclass[a4paper]{jpconf}
\usepackage{color} 
\usepackage{xcolor} 
\usepackage{relsize} 
\usepackage{dsfont}
\usepackage{bbold}
\usepackage{amssymb}
\usepackage{amsmath}
\usepackage{graphicx}
\usepackage[cmtip,arrow]{xy}
\usepackage{pb-diagram,pb-xy}

\def\cmath{\color{blue}}
\def\ctxt{\color{black}}
\DeclareMathOperator*{\argmax}{arg\,max}

\newcommand{\barket}[1]{\left|#1\right\rangle} 
\newcommand{\brabar}[1]{\left\langle#1\right|} 
\newcommand{\braket}[1]{\left\langle#1\right\rangle} 
\def\bmat{\begin{pmatrix}}
\def\emat{\end{pmatrix}}
\def\C{\mathbb{C}} 
\def\N{\mathbb{N}} 
\def\Q{\mathbb{Q}} 
\def\R{\mathbb{R}} 
\def\Z{\mathbb{Z}}
\newcommand{\cabs}[1]{\left|#1\right|} 
\def\denmat{\mathlarger{\rho}}
\newcommand{\denset}[2]{\mathcal{D}_{#1}\farg{#2}} 
\DeclareMathOperator{\diag}{diag}
\ifx\e\undefined\DeclareMathOperator{\e}{e}\fi
\def\energyp{\boldsymbol{\varepsilon}}
\def\Entropy{\mathrm{\mathbf{S}}} 
\newcommand{\farg}[1]{\!\left(#1\right)} 
\def\Hspace{\mathcal{H}} 
\def\imu{\mathrm{i}} 
\DeclareMathOperator{\idmat}{\mathds{1}}

\newcommand{\inner}[2]{\left\langle#1\mid#2\right\rangle} 
\newcommand{\innerform}[3]{\braket{#1\cabs{#2}#3}} 
\newcommand{\ketbra}[1]{\barket{#1}\!\brabar{#1}} 
\DeclareMathOperator*{\KroneckerProduct}{\text{\raisebox{0.3ex}{$\mathsmaller{\bigotimes}$}}}
\def\lcycle{\boldsymbol{\ell}}	
\newcommand{\Math}[1]{$\cmath{}#1$} 
\newcommand{\MathEq}[1]{\begin{equation*}\cmath{#1}\end{equation*}}
\newcommand{\MathEqLab}[2]{\begin{equation}\cmath{#1}\label{#2}\end{equation}}

\newcommand{\MathEqArrLab}[1]{\begin{align}\cmath{}#1\end{align}}
\newcommand{\norm}[1]{\left\lVert#1\right\rVert}
\DeclareMathOperator{\ord}{ord} 
\newcommand{\ordset}[1]{\left[#1\right]} 
\def\Prob{\mathrm{\mathbf{P}}} 
\newcommand{\projector}[1]{\Pi_{#1}}
\def\repq{\mathsf{U}} 
\def\runisymb{\mathsf{r}} 
\def\rcycle{\mathit{m}}	
\newcommand{\runi}[1]{\runisymb_{#1}} 
\newcommand{\set}[1]{\left\{#1\right\}} 
\DeclareMathOperator{\spn}{span}
\newcommand{\SymG}[1]{\mathsf{S}_{#1}} 
\newcommand{\timeder}[1]{\,\dot{#1}} 
\ifx\tr\undefined\DeclareMathOperator{\tr}{tr}\fi

\def\U{\mathsf{U}} 
\newcommand{\UG}[1]{\U\!\vect{#1}} 
\newcommand{\vect}[1]{\left(#1\right)} 
\newcommand{\Vthree}[3]{\bmat#1\\#2\\#3\emat} 
\def\wG{\mathsf{G}} 
\def\wS{\Omega} 
\def\wSN{\mathsf{N}} 

\begin{document}
\title{Quantum models based on finite groups}
\author{Vladimir V Kornyak}
\address{Laboratory of Information Technologies,
           Joint Institute for Nuclear Research\\
           141980 Dubna, Russia}
\ead{vkornyak@gmail.com}
\begin{abstract}%
We consider a constructive modification of quantum-mechanical formalism.
Replacement of a general unitary group by unitary representations of finite groups makes it possible to reproduce quantum formalism without loss of its empirical content.
Since any linear representation of a finite group can be implemented as a subrepresentation of a permutation representation, quantum-mechanical problems can be formulated in terms of permutation groups.
Reproducing quantum behavior in the framework of permutation representations of finite groups makes it possible to clarify the meaning of a number of physical concepts.
\end{abstract}
\section{Introduction}\label{intro}
Physics, being an empirical science, is insensitive to the replacement of infinite concepts in physical theories by finite ones.
The reason for such a replacement is that effective modeling is possible only if the problem is formulated in finite terms.
Moreover, it is preferable to formulate physical problems in finite terms from the very beginning, since infinities can lead to both descriptive losses and artifacts.
In this context, Paul Dirac made the often quoted statement that the most important challenge in physics is ``\emph{to get rid of infinity}''.
In this paper, we describe a constructive version of quantum formalism that does not involve any concepts associated with actual infinities.
\par
The main part of the paper begins with a summary of the basic concepts of the standard quantum mechanics with emphasis on the aspects important for our purposes.
\par
Further we describe a constructive modification of the quantum formalism.
We start with replacing a unitary group --- a continuous group of symmetries of quantum states --- by a finite group.
The natural consequence of this replacement is unitarity: any linear representation of a finite group is unitary.
Any finite group is naturally associated with some cyclotomic field --- an extension of the rational numbers by roots of unity.
Generally, a cyclotomic field is a dense subfield of the field of complex numbers. 
This can be regarded as an explanation of the presence of complex numbers in the quantum formalism.
Any linear representation of a finite group over a cyclotomic field associated with the group can be obtained from a permutation action of the group on vectors with natural components by projection into suitable invariant subspace.
This allows us to reproduce all the elements of quantum formalism in invariant subspaces of permutation representations.
\par
We consider a model of quantum evolution inspired by the quantum Zeno effect%
\footnote{The \emph{quantum Zeno effect} (also known as the \emph{Turing paradox}) is an experimentally verifiable property of a quantum system to follow a prescribed evolution when a properly prepared sequence of observations is applied.}
--- the most convincing manifestation of the role of observation in the dynamics of quantum systems.
The model represents the quantum evolution as a sequence of observations with unitary transitions between them.
Standard quantum mechanics assumes a single deterministic unitary transition between observations.
We generalize this assumption. We treat a unitary transition as a kind of gauge connection --- a way of identifying indistinguishable entities at different times.
\emph{A priori}, any unitary transformation can be used as a data identification rule.
So, we assume that all unitary transformations participate in transitions between observations with appropriate weights.
We call a unitary evolution \emph{dominant} if it provides the maximum transition probability.%
\footnote{The principle of least action in physical theories actually implies the selection of dominant evolutions among all possible (``virtual'') evolutions.
The apparent determinism of these selected evolutions can be explained by the sharpness of their dominance.}
The Monte Carlo simulation shows a sharp dominance of such evolutions over other evolutions.
We also present the Lagrangian of the continuum approximation of the model in order to illustrate the typical assumptions made in the transition to a continuous description.
\section{Basic concepts of quantum mechanics}\label{secQM}
Here is a brief outline of the basic concepts of quantum mechanics.
We divide these concepts into three categories:
{\emph{states}}, {\emph{observations and measurements}}, and {\emph{time evolution}}.
\subsection{Quantum states}\label{subsecS}
\paragraph{Pure states:}
A \emph{pure quantum state} is a ray in a Hilbert space \Math{\Hspace} over the field of complex numbers \Math{\C}.
A ray is an \emph{equivalence class} of vectors \Math{\barket{\psi}\in\Hspace} with respect to the \emph{equivalence relation} \Math{\barket{\psi}\sim{}a\barket{\psi},~ a\in\C\setminus\set{0}}.
\par
We can reduce the equivalence classes by the \emph{normalization}: \Math{\barket{\psi}\sim{}\e^{\imu\alpha}\barket{\psi}, \norm{\psi}=1, \alpha\in\R}.
\par
The phase ``degree of freedom'' \Math{\alpha} can be eliminated by transition to the rank one \emph{projector} \Math{\projector{\psi}=\ketbra{\psi}},
which is a special case of a \emph{density matrix}.
\paragraph{Mixed states:}
A \emph{mixed quantum state}  is described by a \emph{general density matrix} \Math{\denmat} characterized by the properties: 
(a)~\Math{\denmat=\denmat^\dagger},  (b)~\Math{\innerform{\psi}{\denmat}{\psi}\geq0} for any \Math{\barket{\psi}\in\Hspace},  (c)~\Math{\tr\denmat=1}.
The set of all density matrices will be denoted by \Math{\denset{}{\Hspace}}.
\par
Any mixed state is a weighted mixture of pure states, that is, any density matrix can be represented in the form \Math{\denmat=\sum_k{p_k}\projector{e_k}},
where \Math{\set{p_k}} is a probability distribution in the ensemble of pure states \Math{\projector{e_k}=\ketbra{e_k}}, \Math{\set{e_k}} is an orthonormal basis in \Math{\Hspace}.
\par
On the other hand, any mixed state can be obtained from a pure state in a larger Hilbert space by taking a partial trace.
This is called \emph{`purification'} and can be done as follows.
Let \Math{\denmat\cong\diag\farg{p_1,p_2,\ldots,p_n}} be a density matrix in a Hilbert space \Math{\Hspace_X,\, \dim\Hspace_X=n}.
If \Math{\Hspace_Y} is another \Math{n}-dimensional Hilbert space, then \Math{\denmat=\tr_Y\vect{\ketbra{\psi}}}, 
where \Math{\barket{\psi}=\sum_k\sqrt{p_k}\barket{e_k}\KroneckerProduct\barket{f_k}} is a pure state in the Hilbert space \Math{\Hspace_X\KroneckerProduct\Hspace_Y}, 
\Math{\tr_Y} denotes the trace over \Math{\Hspace_Y}, \Math{\set{e_k}} and \Math{\set{f_k}} are orthonormal bases in \Math{\Hspace_X} and \Math{\Hspace_Y}, respectively.
\paragraph{States of composite system:}
The Hilbert space of a \emph{composite system} is the tensor product of Hilbert spaces of the components:  
\Math{\Hspace=\Hspace_{1}{\KroneckerProduct}\Hspace_{2}{\KroneckerProduct}\cdots}.
The \emph{states of composite system} are classified into two types: \emph{separable} and \emph{entangled} states.
\par
The set of \emph{\textbf{separable}} states, \Math{\denset{\mathrm{S}}{\Hspace}}, consists of the states \Math{\denmat\in\denset{}{\Hspace}} 
that can be represented as weighted sums of the tensor products of states of the constituents:
\MathEq{\denmat=\sum_{k}w_k\farg{\denmat_{1}^k{\mathsmaller{\KroneckerProduct}}\denmat_{2}^k{\mathsmaller{\KroneckerProduct}}\cdots}\!,
\,w_k\geq0,\,\sum_{k}w_k=1,\,\denmat_{1}^k\in\denset{}{\Hspace_{1}},\,\denmat_{2}^k\in\denset{}{\Hspace_{2}},\,\ldots\,.}
\par
The set of \emph{\textbf{entangled}} states, \Math{\denset{\mathrm{E}}{\Hspace}}, 
is by definition the complement of \Math{\denset{\mathrm{S}}{\Hspace}} in the set of all states:
\Math{\denset{\mathrm{E}}{\Hspace}=\denset{}{\Hspace}\setminus\denset{\mathrm{S}}{\Hspace}}.
\subsection{Observations and measurements}\label{subsecOM}
The terms `observation' and `measurement' are often regarded as being synonymous.
However, it makes sense to separate these concepts.
We treat observation as a more general concept which does not imply, in contrast to measurement, obtaining numerical information. 
\paragraph{Observation} is the detection (``click of detector'') of a system, that is in the state \Math{\denmat}, in the subspace \Math{\mathcal{S}\leq\Hspace}.
The mathematical abstraction of the ``detector in the subspace'' \Math{\mathcal{S}} of a Hilbert space is the operator of projection, \Math{\projector{\mathcal{S}}}, into this subspace.
The result of quantum observation is random and its statistics is described by a probability measure defined on subspaces of the Hilbert space.
Any such legitimate measure \Math{\mu\farg{\cdot}} must be additive on any set of mutually orthogonal subspaces of a Hilbert space: if, e.g., \Math{\mathcal{A}} and \Math{\mathcal{B}}
are mutually orthogonal subspaces, then \Math{\mu\farg{\spn\farg{\mathcal{A}, \mathcal{B}}}=\mu\farg{\mathcal{A}}+\mu\farg{\mathcal{B}}}.
Gleason proved \cite{Gleason} that, excepting the case \Math{\dim\Hspace=2}, 
the only such measures have the form \Math{\mu_\denmat\farg{\mathcal{S}}=\tr\farg{\denmat\projector{\mathcal{S}}}}, where \Math{\denmat} is an arbitrary density matrix.%
\footnote{In fact, Gleason's theorem establishes a one-to-one correspondence between measures and quantum states in the case \Math{\dim\Hspace\neq2}.
In the two-dimensional case (the case of a qubit) the set of admissible  measures is ``greater'' than the set of states, but the measures that are not associated with quantum states are considered non-physical.}
If, in particular, \Math{\denmat} describes a pure state, \Math{\denmat=\ketbra{\psi}}, and \Math{\mathcal{S}} is one-dimensional, \Math{\mathcal{S}=\spn\farg{\barket{\varphi}}}, we come to the familiar Born rule: \Math{\tr\farg{\denmat\projector{\mathcal{S}}}=\Prob_{\text{Born}}=\cabs{\inner{\varphi}{\psi}}^2}.
\paragraph{Measurement} is a special case of observation, 
when the partition of a Hilbert space into mutually orthogonal subspaces is specified by a Hermitian operator \Math{A}, called ``observable''.
Let \Math{\set{e_k}} be an \emph{orthonormal basis of eigenvectors} of \Math{A} and let \Math{\set{a_k}} be the \emph{spectrum} of \Math{A}.
In the basis \Math{\set{e_k}} the operator is written as \Math{A=\sum_ka_k\projector{e_k}}, where \Math{\projector{e_k}=\ketbra{e_k}}. 
``Click of the detector'' \Math{\projector{e_k}} is interpreted as that the \emph{eigenvalue} \Math{a_k\in\R} is the result of the measurement.
The mean for multiple measurements tends to the \emph{expectation value} of \Math{A} in the state \Math{\denmat}: 
\Math{\left\langle{A}\right\rangle_{\!\denmat}=\tr\farg{\denmat{A}}}.
\par
Note that \emph{observation} can formally be treated as a \emph{measurement} if we regard the  projector \Math{\projector{\mathcal{S}}} as an observable
with eigenvalues belonging to the Boolean domain \Math{\set{0,1}} (or \Math{\set{\mathrm{no,yes}}}).
\subsection{Time evolution}\label{subsecTE}
%
Observation and measurement are timeless concepts in the standard quantum formalism.
Time appears as a parameter in describing the unitary transformation of data between observations.
The unitary evolution of the density matrix has the form 
\MathEqLab{\denmat_{t'}=U_{t't}\denmat_tU_{t't}^{\dagger},}{EvoRho}
where \Math{\denmat_{t}} is the state \emph{after} observation (sometimes called in this context ``\emph{preparation}'') at the time \Math{t},
 \Math{\denmat_{t'}} is the state \emph{before} observation at the time \Math{t'}, and \Math{U_{t't}} is the unitary transition between the observations.
In standard quantum formalism, time is considered as a continuous parameter, and relation \eqref{EvoRho} becomes the \emph{von Neumann equation} in the infinitesimal limit.
The evolution of a pure state can be written as \MathEq{\barket{\psi_{t'}}=U_{t't}\barket{\psi_t},} and the corresponding infinitesimal limit is the \emph{Schr\"odinger equation}.
The operator of unitary evolution can be represented as
\MathEqLab{U_{t't}=\e^{2\pi\imu{H}\Delta{t}},}{UfromH}
where \Math{H} is a Hermitian operator called the \emph{Hamiltonian}, \Math{\Delta{t}=t'-t}.
The concept of a Hamiltonian is especially useful in the case of continuous time.
In the discrete time \Math{\Delta{t}} is a natural number, \Math{\Delta{t}\in\N}, and instead of \eqref{UfromH} we can write \MathEq{U_{t't}=U_{\!1}^{\Delta{t}},} where  \Math{U_{\!1}} is the operator of evolution for a unit time step.
\par
To emphasize the role of observation in quantum physics, we accentuate that unitary evolution is simply a change of coordinates in Hilbert space, and it alone is insufficient to describe observable physical phenomena. 
\subsection{Emergence of geometry in large Hilbert space due to entanglement}\label{subsecEG}
Quantum-mechanical theory does not need a geometric space as a fundamental concept --- everything can be formulated using only the Hilbert space formalism.
In this view, the observed geometry must emerge as an approximation.
The currently popular idea \cite{Raamsdonk,MalSus,Cao} of the emergence of geometry within a Hilbert space is based on the notion of entanglement.
Briefly, the scheme of extracting geometric manifold from the entanglement structure of a quantum state \Math{\denmat} in a Hilbert space \Math{\Hspace} is as follows:
\begin{itemize}
	\item
The Hilbert space decomposes into a large number of tensor factors:  \Math{\Hspace=\KroneckerProduct_x\Hspace_x,~ x\in{X}}.
Each factor is treated as a point (or bulk) of geometric space to be built.
A graph \Math{G} --- called \emph{tensor network} ---  with vertices \Math{x\in{X}} and edges \Math{\vect{x,y}\in{}X\times{}X} is introduced.
	\item
The edges of \Math{G} are assigned \emph{weights} based on a \emph{measure of entanglement}, a function that vanishes on separable states and is positive on entangled states.
A typical such measure is the \emph{mutual information}:
\Math{\mathlarger{I}\!\vect{\denmat_{xy}}=S\!\vect{\denmat_{x}}+S\!\vect{\denmat_{y}}-S\!\vect{\denmat_{xy}}},~ where~\Math{\denmat_{x}} denotes the result of taking traces of \Math{\denmat} over all tensor factors excepting the \Math{x}-th (and similarly for \Math{\denmat_{y}} and \Math{\denmat_{xy}});~ \Math{S\!\vect{a}=-\tr\vect{a\log{a}}} is the \emph{von Neumann entropy}.
The graph \Math{G} is supplied with a metric derived from the weights of the edges.
	\item
Finally, the graph \Math{G} is approximately isometrically embedded in a smooth metric manifold of as small as possible dimension using algorithms like \emph{multidimensional scaling} (MDS).
\end{itemize}
\section{Constructive modification of quantum formalism}\label{secCoQM}
David Hilbert, a prominent advocate of the free use of the concept of infinity in mathematics,
wrote the following about the relation of the infinite to the reality:
``\emph{Our principal result is that the {infinite} is nowhere to be found in reality.
It {neither exists in nature} {nor provides} a legitimate {basis for rational thought} --- a remarkable harmony between being and thought.}''
Adopting this view, we can not but conclude that all the concepts of quantum mechanics can be formulated in constructive finite terms without distorting the empirical content \cite{Kornyak13,Kornyak15,Kornyak16}.
\subsection{Losses due to continuum and differential calculus}\label{subsecDIFF}
Differential calculus --- including differential equations, differential geometry, etc. --- underlies  mathematical methods in physics.
The applicability of differential calculus is based on the assumption that any relevant function can be approximated by linear relations on a small scale.
This assumption simplifies many problems in physics and mathematics, but at the cost of loss of completeness.
\par
As an example, consider the problem of classifying simple groups.
The concept of a group is an abstraction of the properties of \emph{permutations} (also called \emph{one-to-one mappings} or \emph{bijections}) of a set.
Namely, an abstract group is a set with an \emph{associative} operation, an \emph{identity} element, and an \emph{invertibility} for each element.
Additional assumptions allow to make the concept of a group more meaningful.
The most natural of these assumptions is: (a) \emph{the group is finite}.
It is clear that empirical physics is insensitive to this assumption: ultimately, any empirical description is reduced to a finite set of data.
A simple complement to assumption (a) --- \emph{the group is infinite} --- is of little use for a number of reasons, e.g.,
there are one-to-one mappings between infinite sets and their proper subsets.
Thus, more restrictive assumptions are used, the most important of which is: (b) \emph{the group is a differentiable manifold} --- such a group is called \emph{Lie group}.
Assumption (b) imposes severe constraints on possible physical models.
\par
The problem of classification of simple groups%
\footnote{\emph{Simple groups}, that is, not containing non-trivial \emph{normal subgroups}, are ``building blocks'' for all other groups.}
under assumption (b) turned out to be rather easy and was solved by two people (Killing and Cartan) in a few years.
The result is four infinite series: \Math{A_n}, \Math{B_n}, \Math{C_n}, \Math{D_n}; and five exceptional groups: \Math{E_6}, \Math{E_7}, \Math{E_8}, \Math{F_4}, \Math{G_2}.
\par
The solution of the classification problem under assumption (a) required the efforts of about a hundred people for over a hundred years \cite{Solomon}.
But the result --- \emph{``the enormous theorem''} --- is much richer.
The list of finite simple groups contains \Math{16+1+1} infinite series:
\begin{itemize}
	\item \emph{groups of Lie type}: \\
\Math{A_n(q)}, \Math{B_n(q)}, \Math{C_n(q)}, \Math{D_n(q)}, \Math{E_6(q)}, \Math{E_7(q)}, \Math{E_8(q)}, \Math{F_4(q)}, \Math{G_2(q)},\\[2pt]
\Math{^2A_n\vect{q^2}}, \Math{^2B_n\vect{2^{2n+1}}}, \Math{^2D_n\vect{q^2}}, \Math{^3D_4\vect{q^3}},
\Math{^2E_6\vect{q^2}}, \Math{^2F_4\vect{2^{2n+1}}}, \Math{^2G_2\vect{3^{2n+1}}};
	\item \emph{cyclic groups of prime order},~ \Math{\Z_p};
	\item \emph{alternating groups},~ \Math{A_n,~n\geq5};
\end{itemize}
and \Math{26} \emph{sporadic groups}:
\Math{M_{11}}, \Math{M_{12}}, \Math{M_{22}}, \Math{M_{23}}, \Math{M_{24}}, \Math{J_1}, \Math{J_2}, \Math{J_3}, \Math{J_4}, \Math{Co_1}, \Math{Co_2}, \Math{Co_3},\\
\phantom{and \Math{26} \emph{sporadic groups}~\,:}\Math{Fi_{22}}, \Math{Fi_{23}}, \Math{Fi_{24}}, \Math{HS}, \Math{McL}, \Math{He}, \Math{Ru}, \Math{Suz}, \Math{O'N}, \Math{HN},
\Math{Ly}, \Math{Th}, \Math{B}, \Math{M}.\\
Note that finite groups have an advantage over Lie groups in the sense that in empirical applications any Lie group can be modeled by some finite group, but not vice versa. 
\subsection{Replacing unitary group by finite group}\label{subsecRemInf}
The main non-constructive element of the standard quantum formalism is the unitary group \Math{\UG{n}}, a set of cardinality of the continuum.
\par
Formally, the group \Math{\UG{n}} can be replaced by some finite group which is empirically equivalent to \Math{\UG{n}} as follows.
From the theory of quantum computing it is known that \Math{\UG{n}} contains a dense finitely generated --- and, hence, countable --- matrix subgroup \Math{\U_*\!\vect{n}}.
The group \Math{\U_*\!\vect{n}} is \emph{residually finite} \cite{Malcev}, i.e. it has a reach set of non-trivial homomorphisms to finite groups \cite{Magnus}.
\par
In essence, instead of deriving a finite group from an infinite group, it is more natural to assume that the fundamental symmetries are presented by finite groups, 
and \Math{\UG{n}}'s are just continuum approximations of their unitary representations.
\par
The following properties of finite groups are important for our purposes:
\begin{itemize}
	\item
any finite group is a subgroup of a \emph{symmetric group},
\item
any linear representation of a finite group is \emph{unitary},
\item
any linear representation is  subrepresentation of some \emph{permutation representation}.
\end{itemize}
\subsection{``Physical'' numbers and finite groups}\label{subsecPhysNumb}
The basic number system in quantum formalism is the complex field \Math{\C}.
This non-constructive field can be obtained as a metric completion of many algebraic extensions of rational numbers.
We consider here constructive numbers that are closely related to finite groups and are based on two primitives with a clear intuitive meaning:
\begin{enumerate}
	\item
\emph{natural numbers} (``counters''): \Math{\N=\set{0,1,\ldots}};
	\item
\Math{k}th \emph{roots of unity}%
\footnote{There are \Math{k} different \Math{k}th roots of unity.
A \Math{k}th root of unity is called \emph{primitive} if \Math{\runi{k}^m=1} only for \Math{m\equiv0\mod{k}}.}
 (``algebraic form of the idea of \Math{k}-periodicity''): \Math{\runi{k}\mid\runi{k}^k=1}.
\end{enumerate}
These basic concepts are sufficient to represent all physically meaningful numbers.
\par
We start by introducing \Math{\N\!\ordset{\runi{k}}}, the extension of the \emph{semiring} \Math{\N} by primitive \Math{k}th root of unity.
The extension \Math{\N\!\ordset{\runi{k}}} is a \emph{ring} if \Math{k\geq2}.
This construction allows, in particular, to add \emph{negative numbers} to the naturals --- 
the ring of integer numbers is the extension of \Math{\N} by the primitive square root of unity: \Math{\Z=\N\!\ordset{\runi{2}}}.
Further, by a standard mathematical procedure (applicable to any \emph{integral domain} --- a commutative ring without zero divisors), we construct the \Math{k}th \emph{cyclotomic field} as the \emph{fraction field} of the ring \Math{\N\!\ordset{\runi{k}}}:
\Math{\Q\farg{\runi{k}}=\mathrm{Frac}\farg{\N\!\ordset{\runi{k}}}}.
For \Math{k=2} we have the field of rational numbers: \Math{\Q=\mathrm{Frac}\farg{\N\!\ordset{\runi{2}}}}.
If \Math{k\geq3}, then the field \Math{\Q\!\vect{\runi{k}}} is a \emph{dense subfield} of \Math{\C}, i.e. (constructive) cyclotomic fields are empirically indistinguishable from the (non-constructive) complex field.
\par
The importance of cyclotomic numbers for constructive quantum mechanics is explained by the following.
Let us recall some terms.
The \emph{exponent} of a group \Math{\wG} is the least common multiple of the orders of its elements.
A \emph{splitting field} for a group \Math{\wG} is a field that allows to split completely any linear representation of \Math{\wG} into irreducible components.
A \emph{minimal splitting field} is a splitting field that does not contain proper splitting subfields.
Although minimal splitting field for a given group \Math{\wG} may be non-unique, any minimal splitting field is a subfield of some cyclotomic field \Math{\Q\farg{\runi{k}}}, 
where \Math{k} is a divisor of the exponent of \Math{\wG}.
If we want to minimize computations, we can select the field \Math{\Q\farg{\runi{k}}} with the smallest \Math{k} among cyclotomic splitting fields.
Thus, to work with any representation of \Math{\wG} it is sufficient to use the \Math{k}th cyclotomic field, where \Math{k} is determined by the structure of \Math{\wG}.
\par
An important field associated with a group is a so-called `sufficiently large field'.
A \emph{sufficiently large field} for \Math{\wG} is defined as a field that contains all the \Math{m}th roots of unity, where \Math{m} is the exponent of \Math{\wG}.
In particular, the \Math{m}th cyclotomic field  \Math{\Q\farg{\runi{m}}} is a sufficiently large field.
A {sufficiently large field} for \Math{\wG} is a splitting field for \emph{every subgroup} of \Math{\wG}.
This fact is useful in considering a symmetry breaking, which is actually a transition to a subgroup.
\subsection{Constructive representations of a finite group}\label{subsecConstrRep}
\begin{figure}[ht]
\centering
\includegraphics[width=0.7\textwidth]{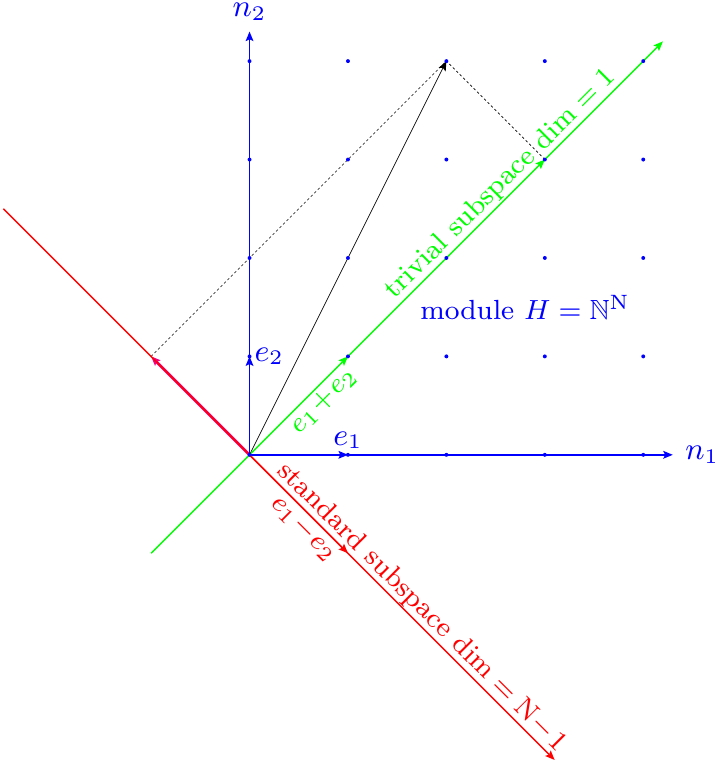}
\caption{Decomposition of the natural representation of \Math{\SymG{{\wSN}}} into irreducible components}
\label{NatStdS2}
\end{figure}
Let a group \Math{\wG} act by permutations on a set \Math{\wS,~ \cabs{\wS}=\wSN}.
If we assume that the elements of \Math{\wS} are ``types'' of some discrete entities ({``ontological entities'', ``elements of reality''}), 
then collections of these entities can be described as elements of the module  \Math{H=\N^\wSN} with the basis \Math{\wS} over the semiring of natural numbers \Math{\N}. 
The decomposition of the action of \Math{\wG} in the module \Math{H} into irreducible components reflects the structure of the invariants of the action.
In order for the decomposition to be complete, it is necessary to extend the semiring \Math{\N} to a splitting field, 
e.g., to a cyclotomic field \Math{\Q\farg{\runi{k}}}, where \Math{k} is a suitable divisor of the exponent of \Math{\wG}. 
With such an extension of the scalars, the module \Math{H} is transformed into a Hilbert space \Math{\Hspace} over \Math{\Q\farg{\runi{k}}}.
This construction, with a suitable choice of the permutation domain \Math{\wS}, 
allows us to obtain \emph{any representation} of the group \Math{\wG} in some invariant subspace of the Hilbert space \Math{\Hspace}.
We obtain ``quantum mechanics'' within an invariant subspace if, in addition to unitary evolutions, projective measurements are also restricted by this subspace.
\par
The above is illustrated in figure \ref{NatStdS2} by the example of the natural action of the symmetric group \Math{\SymG{{\wSN}}} on the set
\Math{\wS=\set{e_1,\ldots,e_{\wSN}}}.
The \emph{natural} representation of \Math{\SymG{{\wSN}}} decomposes into two irreducible components: one-dimensional \emph{trivial} and \Math{\vect{\wSN-1}}-dimensional \emph{standard} representations.
The spaces of these representations have, respectively, the following canonical bases: \Math{e_1+e_2+\cdots+e_{\wSN}} and \Math{e_1-e_2, e_2-e_3,\ldots,e_{\wSN-1}-e_{\wSN}}.
Note, that any symmetric group is a so-called \emph{rational-representation} group, 
i.e. the field of rational numbers \Math{\Q} is a splitting field for \Math{\SymG{{\wSN}}}. 
%
\section{Modeling quantum evolution}\label{secModel}
\subsection{Time} 
For description of evolution, we use a discrete model of time.
We define the fundamental (``Planck'') time as the ordered sequence of natural numbers:
\Math{\mathcal{T}=\N}.
We introduce the ``empirical time'' as a finite sequence of ``instants of observations'':
\MathEqLab{\mathrm{T}=\ordset{t_{0},t_{1},\ldots,t_{k-1},t_{k},\ldots,t_{\mathrm{n}}}.}{Tobserv}
The simplest assumption is that  \Math{\mathrm{T}} is a subsequence of \Math{\mathcal{T}}.
A more realistic model (which we will not develop here) could be a sequence of distributions%
\footnote{The smallest uncertainty in fixing time instants available in modern physics is about \Math{10^{26}} Planck units.}
around points \Math{t_k\in\mathcal{T}}, like, e.g., the \emph{binomial distribution}
\MathEq{K_{\sigma}\farg{\tau-t_k}=\frac{\vect{2\sigma}!}{4^{\sigma}\vect{\sigma-t_k+\tau}!\vect{\sigma+t_k-\tau}!},
\hspace{10pt}t_k-\sigma\leq\tau\leq{t_k}+\sigma.}
%
\subsection{Model of quantum evolution} 
The data of the model include the sequence of the length \Math{\mathrm{n}+1} for states
\MathEqLab{\ordset{\denmat_{0},\denmat_{1},\ldots,\denmat_{k-1},\denmat_{k},\ldots,\denmat_{\mathrm{n}}}}{seqStates}
and the sequence of the length \Math{\mathrm{n}} for unitary transitions between observations
\MathEqLab{\ordset{U_{1},\ldots,U_{k},\ldots,U_{\mathrm{n}}}.}{seqU}
Standard quantum mechanics presupposes a single unitary evolution, \Math{U_k}, between observations at times \Math{t_{k-1}} and \Math{t_k}.
The \emph{single-step transition probability} takes the form
\MathEq{\Prob_k=\tr\farg{U_k\denmat_{k-1}U_k^\dagger\denmat_k}.} 
The evolution can be expressed in terms of the Hamiltonian \Math{H_k}: \Math{U_{k}=\e^{-2\pi\imu{H_k}\vect{t_k-t_{k-1}}}}.
In physical theories, Hamiltonians are usually derived from \emph{the principle of least action}.
Like any extremal principle, it implies the selection of a small subset of the dominant elements in a large set of candidates.
Thus it is natural to assume that any unitary evolution takes part in the transition between observations with appropriate weight, but only the dominant evolutions are manifested in observations.
Therefore, in our model, we use the following modification of the single-step transition probability
\MathEq{\Prob_k=\sum_{g\in\wG}w_{kg}\tr\farg{\repq\farg{g}\denmat_{k-1}\repq\farg{g^{-1}}\denmat_k},} 
where \Math{\wG} is a finite group,  
\Math{\repq} is a  unitary representation of \Math{\wG},  
\Math{w_{kg}} is the weight of \Math{g\in\wG} at \Math{k}th transition: \Math{w_{kg}\geq0,~\sum_{g\in\wG}w_{kg}=1}.
\par
The \emph{single-step entropy} is defined as \MathEqLab{\Delta\Entropy_{k}\!=-\log\Prob_{k}.}{Entropy1}
Continuum approximation of \eqref{Entropy1} leads to the \emph{Lagrangian} \Math{\mathcal{L}}.
\par
The probability of the whole trajectory is
\MathEqLab{\Prob_{0\rightarrow{\mathrm{n}}}=\prod_{k=1}^\mathrm{n}\Prob_k.}{Ptraj}
Taking the logarithm of \eqref{Ptraj}, we arrive at the \emph{entropy of trajectory} 
\MathEq{\Entropy_{{0}\rightarrow{\mathrm{n}}}\!=\sum_{k=1}^\mathrm{n}\Delta\Entropy_{k},} 
the continuum approximation of which is the \emph{action} \Math{\mathcal{S}=\int\!\mathcal{L}dt}. 
\subsection{Continuum approximation of discrete model}\label{subsecDiscrCont}
%
To build a continuum approximation, we immerse sequence of observation times \eqref{Tobserv} in the continuous time interval \Math{\ordset{t_0,t_{\mathrm{n}}}\subseteq\R},
replace sequences \eqref{seqStates} and \eqref{seqU} with the continuous differentiable functions \Math{\denmat\farg{t}} and \Math{U\farg{t}},
and go to the limit \Math{\mathrm{n}\rightarrow\infty,~\Delta{t}_k=t_k-t_{k-1}\rightarrow0}.
Using the Taylor expansion of \Math{\denmat\farg{t}} and \Math{U\farg{t}} at the point \Math{t_{k-1}}, we can compute the approximation for the single-step transition probability:
\MathEq{\Prob_k\approx{}P_k^{(0)}-P_k^{(1)}\Delta{t}_k-P_k^{(2)}\Delta{t}_k^2-\cdots\,.} 
We see that product \eqref{Ptraj} can have a nonzero limit only if \Math{P_k^{(0)}=1} for all but a finite number of factors.
This is an artificial restriction imposed by the introduction of infinity.
The term \Math{P_k^{(0)}} has the meaning of the ``zero-step transition probability''.
Its value is \Math{P_k^{(0)}=\tr\farg{\denmat_{k-1}^2}}. 
Since \Math{\tr\farg{\denmat^2}=1} only for pure states, \Math{\denmat=\barket{\psi}\!\brabar{\psi}}, 
we can not construct a reasonable continuum approximation in the case of general mixed states for which \Math{\tr\farg{\denmat^2}<1}.
Calculation for the case of a pure state shows that
\MathEq{\Prob_k\approx{}1-P_k^{(2)}\Delta{t}_k^2-\cdots\,,} 
and the {single-step entropy} has the following approximation \MathEqLab{\Delta\Entropy_{k}\!=-\log\Prob_{k}\approx{}P_k^{(2)}\Delta{t}_k^2+\cdots\,.}{Entropy1appr}
The leading-order (i.e. the second-order) term in expansion \eqref{Entropy1appr} defines a Lagrangian.
For further calculations, we use the following simplifying assumptions:
\begin{itemize}
	\item 
Assuming that unitary matrix \Math{U}  belongs to a unitary representation of a Lie group, we use the Lie algebra approximation, \Math{U\approx\idmat+\imu{}A}, where \Math{A} is a Hermitian matrix.
Thus, we replace the function \Math{U\farg{t}} by a continuous function \Math{A\farg{t}}, which can be regarded as a gauge field (gauge connection) in the \Math{\vect{0+1}}-dimensional ``space-time''.
	\item 
We introduce derivatives \Math{\timeder{A}\farg{t}} and \Math{\timeder{\psi}\farg{t}} as the coefficients in the linear approximations \Math{\Delta{A}_k\approx\timeder{A}\Delta{t}_k} and \Math{\Delta{\psi}_k\approx\timeder{\psi}\Delta{t}_k}, 
where \Math{\Delta{A}_k=A_k-A_{k-1}} and \Math{\Delta{\psi}_k=\psi_k-\psi_{k-1}}.
The time derivative \Math{\timeder{A}\farg{t}} is an analog of the gauge curvature in  \Math{\vect{0+1}} dimensions.
\end{itemize}
Applying the above assumptions and approximations to single-step entropy \eqref{Entropy1appr} and taking the infinitesimal limit, we obtain the Lagrangian:
\MathEqArrLab{\mathcal{L}=&\cmath\underbrace{\innerform{\psi}{\timeder{A}^2}{\psi}-\innerform{\psi}{\timeder{A}}{\psi}^2\!}_{\text{\small\ctxt{}dispersion of \Math{\timeder{A}} in state \Math{\psi}}}
\nonumber\\&\cmath
-\imu\bigg(\innerform{\timeder{\psi}}{\timeder{A}}{\psi}-\innerform{\psi}{\timeder{A}}{\timeder{\psi}}+2\innerform{\psi}{\timeder{A}}{\psi}\inner{\psi}{\timeder{\psi}}\bigg)-\inner{\psi\!}{\!\timeder{\psi}}^2.\label{Lagrangian}}
Note that the first two terms of \eqref{Lagrangian} form the \emph{dispersion} of the ``gauge curvature'' in the state \Math{\psi}:\\
\Math{\farg{\Delta_\psi\!\timeder{A}}^2=\innerform{\psi}{\timeder{A}^2}{\psi}-\innerform{\psi}{\timeder{A}}{\psi}^2},
where the value \Math{\Delta_\psi\!\timeder{A}} is called the \emph{standard deviation}.
\subsection{Dominant evolutions}\label{subsecDomEvol}
Let \Math{\repq} be a unitary representation of a group \Math{\wG} in a Hilbert space \Math{\Hspace}.
For an arbitrary pair of quantum states, \Math{\denmat_1,\denmat_2\in\denset{}{\Hspace}}, we define an element \Math{g_*\in\wG} by
\MathEq{g_*=\argmax\limits_{g\in\wG}\,\tr\farg{\repq\farg{g}\denmat_{1}\repq\farg{g^{-1}}\denmat_2}.}
The operator \Math{\repq\farg{g_*}} maximizes the probability associated with unitary transition between the states \Math{\denmat_1} and \Math{\denmat_2}.
We call such operators \emph{dominant evolutions}.
\par
In the case of natural and standard representations of symmetric groups, the dominant evolutions can be found by a simple algorithm.
Namely, the dominant evolutions from the group \Math{\SymG{\wSN}} between the pure states represented by the vectors from the module \Math{H=\N^\wSN} can be computed as follows.
Let \Math{\barket{n}=\Vthree{n_1}{\vdots}{n_\wSN},~\barket{m}=\Vthree{m_1}{\vdots}{m_\wSN},~\barket{1}=\Vthree{1}{\vdots}{1}}~ be \Math{\wSN}-dimensional vectors with natural components.
The Born probabilities for the pair \Math{\barket{n}} and \Math{\barket{m}} are
\MathEq{\Prob_\mathrm{nat}\vect{\barket{n}\!,\barket{m}}=\cmath\frac{\inner{n}{m}^2}{\inner{n}{n}\inner{m}{m}}}
and
\MathEqLab{\Prob_\mathrm{std}\vect{\barket{n}\!,\barket{m}}=\frac{\vect{\inner{n}{m}-\frac{1}{\wSN}\inner{n}{\!1}\inner{1\!}{m}}^2}
{\vect{\inner{n}{n}-\frac{1}{\wSN}\inner{n}{\!1}\inner{1\!}{n}}\vect{\inner{m}{m}-\frac{1}{\wSN}\inner{m}{\!1}\inner{1\!}{m}}}}{Pstd}
for the natural and standard representations, respectively.
\par
Let \Math{R_a} denote the permutation (as well as its unitary representation), that sorts the components of the vector \Math{\barket{a}} in some (ascending or descending) order.
It is not hard to show that the unitary operator \Math{U=R_m^{-1}R_n} maximizes the probability \Math{\Prob_\mathrm{nat}\farg{U\!\barket{n}\!,\barket{m}}},
if the permutations \Math{R_n} and \Math{R_m} sort the vectors  \Math{\barket{n}} and \Math{\barket{m}} \emph{identically} --- i.e., either both in ascending or both in descending order.
The probability \Math{\Prob_\mathrm{std}\farg{U\!\barket{n}\!,\barket{m}}} is maximized by the operator \Math{U=R_m^{-1}R_n}, 
if \Math{R_n} and \Math{R_m} sort the vectors either \emph{identically} or \emph{oppositely} --- to make selection one should compare the values of the numerator in \eqref{Pstd} for both possibilities (the  denominator is invariant under any permutations).
\subsection{Energy of permutation}\label{subsecPermEn}
Planck's formula, \Math{E={h}\boldsymbol{\nu}}, relates energy to frequency.
This relation is reproduced by the quantum-mechanical definition of energy as an eigenvalue of a Hamiltonian. 
The Hamiltonian associated with a unitary transformation \Math{U} can be written as \Math{H={\imu}{\hbar}\ln{}U}.
\par
Consider the energy spectrum of a unitary operator defined by a permutation.
Let \Math{p} be a permutation of the cycle type 
\Math{\set{{\lcycle_1^{\rcycle_1}},\ldots,{\lcycle_k^{\rcycle_k}},\ldots,{\lcycle_K^{\rcycle_K}}}}, 
where \Math{\lcycle_k} and \Math{\rcycle_k} represent lengths and multiplicities of cycles in the decomposition of \Math{p} into disjoint cycles.
Decomposition into disjoint cycles implies the representation of a permutation by a block diagonal matrix, 
where the block corresponding to the cycle \Math{\vect{1,2,\ldots,\lcycle}} is the \Math{\lcycle\times\lcycle} matrix
\MathEq{C_{\lcycle}=
\bmat
0&1&0&\cdots&0\\
0&0&1&\cdots&0\\
\vdots&\vdots&\vdots&\vdots&\vdots\\
1&0&0&\cdots&0
\emat.}
This matrix can be diagonalized, in particular, by the unitary matrix of the discrete Fourier transform
\MathEq{F_{\lcycle}=\frac{1}{\sqrt{\lcycle}}
\bmat
1&1&1&\cdots&1\\
1&\runi{}&\runi{}^2&\cdots&\runi{}^{\lcycle-1}\\
1&\runi{}^2&\runi{}^4&\cdots&\runi{}^{2\vect{\lcycle-1}}\\
\vdots&\vdots&\vdots&\cdots&\vdots\\
1&\runi{}^{\lcycle-1}&\runi{}^{2\vect{\lcycle-1}}&\cdots&\runi{}^{\vect{\lcycle-1}\vect{\lcycle-1}}
\emat,}
where \Math{\runi{}} is a primitive \Math{\lcycle}th root of unity.
The diagonalized matrix has the form
\MathEq{F^{-1}_{\lcycle}C_{\lcycle}F_{\lcycle}=
\bmat
1&0&0&\cdots&0\\
0&\runi{}&0&\cdots&0\\
0&0&\runi{}^2&\cdots&0\\
\vdots&\vdots&\vdots&\vdots&\vdots\\
0&0&0&\cdots&\runi{}^{\lcycle-1}
\emat.}
After taking the logarithm, we see that the Hamiltonian%
\footnote{We define the Hamiltonian via the relation \Math{U=\e^{2\pi\imu{H}}} to ensure rationality of eigenvalues of \Math{H}.}
 of the permutation \Math{p} has the following diagonal form
 \MathEq{{}H_p=
\bmat
\idmat_{\rcycle_1}\!\otimes\,H_{\lcycle_1}
&&\\
&\hspace{-25pt}
\ddots&\\
&&
\hspace{-25pt}
\idmat_{\rcycle_K}\!\otimes\,H_{\lcycle_K}
\emat,~~~\text{\ctxt{}where}~~~H_{\lcycle_k}={\displaystyle\frac{1}{\lcycle_k}}
\bmat
0&&&\\[-4pt]
&\hspace{-6pt}{1}&&\\[-4pt]
&&\hspace{-6pt}\ddots&\\[-4pt]
&&&\hspace{-8pt}\lcycle_k\!-\!1
\emat.
}
We shall call the least nonzero energy of a permutation the \emph{base energy}:
\MathEq{\displaystyle\energyp
=\frac{1}{\max\vect{\lcycle_1,\ldots,\lcycle_K}}\,.} 
Simulation shows that the base \emph{(``ground state'', ``zero-point'', ``vacuum'')} energy is statistically more significant than other energy levels.
%
\subsection{Monte Carlo simulation of dominant evolutions}\label{subsecMonteCarlo}
\def\hsp{\hspace*{-40pt}}
\def\wid{1.1\textwidth}
\def\hei{0.165\textheight}
\begin{figure}[ht]
\centering
\hsp\includegraphics[width=\wid,height=\hei]{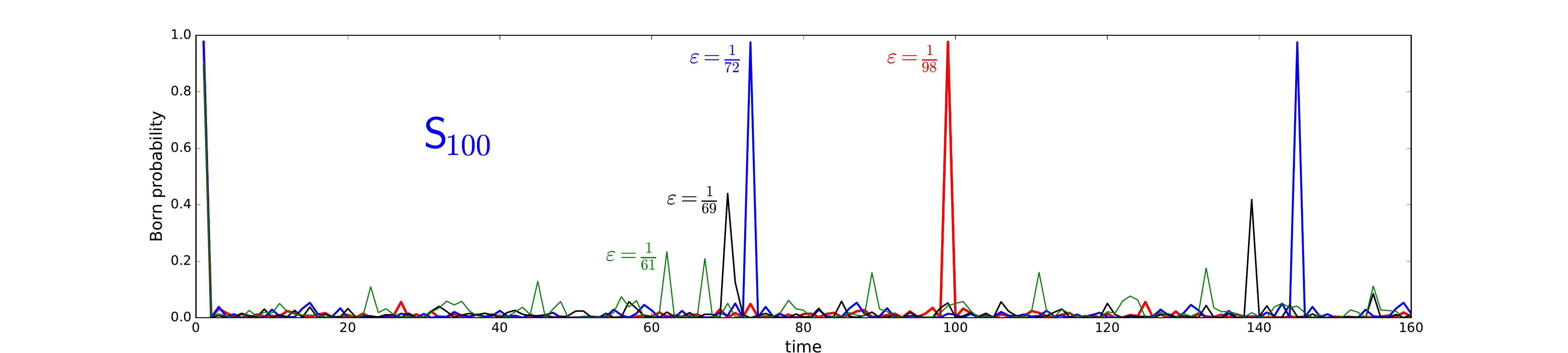}
\hsp\includegraphics[width=\wid,height=\hei]{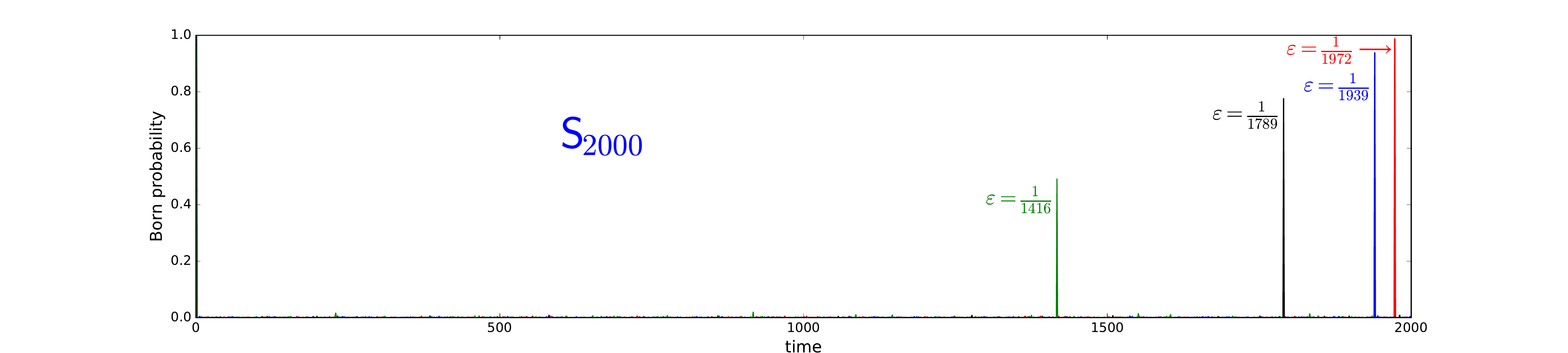}
\caption{Dominant evolutions between randomly generated states. Born probability vs time}
\label{MonteCarlo}
\end{figure}
We can compare the contributions of different evolutions using a Monte Carlo simulation. 
The simulation shows that the contributions of non-dominant evolutions are negligible for sufficiently large groups.
\par
Figure \ref{MonteCarlo} shows several dominant evolutions for the standard representation of the symmetric groups \Math{\SymG{100}} and  \Math{\SymG{2000}}.
Each graph represents the time dependencies of Born's probabilities for the dominant evolutions between four randomly generated pairs of natural vectors.
The dominant evolutions are marked by labeling their peaks with their base energies, \Math{\energyp}, which are listed in table \ref{DomEvoEnOrd}.
The table contains also the orders%
\footnote{The \emph{order} (or \emph{period}) of an element \Math{g} of a group is the smallest integer \Math{m>0} such that \Math{g^m=e}, where \Math{e} denotes the identity of the group.}
of the dominant evolutions. 
\begin{table}[h]
\caption{\label{DomEvoEnOrd}Base energies and orders of dominant evolutions}
\begin{center}
\begin{tabular}{llllllllll}
\br
&\multicolumn{4}{l}{\Math{\SymG{100},~\dim\Hspace=99
}}&\hspace*{10pt}&\multicolumn{4}{l}{\Math{\SymG{2000},~\dim\Hspace=1999
}}\\
\mr
\Math{\energyp}&\Math{\frac{1}{61}}&\Math{\frac{1}{69}}&\Math{\frac{1}{72}}&\Math{\frac{1}{98}}&&\Math{\frac{1}{1416}}&\Math{\frac{1}{1789}}&\Math{\frac{1}{1939}}&\Math{\frac{1}{1972}}\\
\mr
\Math{\ord}&\Math{4026}&\Math{4830}&\Math{72}&\Math{98}&&\Math{3161864216280}&\Math{1061485260}&\Math{93072}&\Math{53244}\\
\br
\end{tabular}
\end{center}
\end{table}
\par
We compared randomly generated evolutions with the dominant ones.
Such non-dominant evolutions are not shown in the figure, since they are almost invisible against the sharp peaks of dominant evolutions.
The superiority of the dominant evolutions increases with the group size.
\section*{Summary}\label{Summary}
\begin{enumerate}
	\item 
The general scheme of quantum mechanics can be completely reproduced in terms of projections of permutation representations of finite groups into invariant subspaces.
	\item 
Quantum randomness is a consequence of the fundamental impossibility of tracing the individuality of indistinguishable entities in their evolution:
the only available objective information about such entities is reduced to the invariants of their symmetry group,
so information about individual events can only be probabilistic.
	\item 
The native number systems for quantum formalism are cyclotomic fields, and the field of complex numbers is just a non-constructive metric completion of cyclotomic fields.	
	\item 
Observable behavior of quantum system is determined by the dominants among all possible evolutions.
	\item 
The principle of least action is a continuum approximation of the principle of selection of the most probable trajectories.	
\end{enumerate}
\ack
I am grateful to Yu.A. Blinkov, V.P. Gerdt and S.I. Vinitsky for fruitful discussions.
\end{document}